\documentclass[conference]{IEEEtran}
\usepackage[utf8]{inputenc}
\usepackage{graphicx}
\usepackage{amsmath}
\usepackage{float}
\usepackage{comment}
\usepackage{amsmath}
\usepackage{booktabs}
\usepackage{graphicx}
\usepackage{amssymb}

\usepackage{siunitx}

\usepackage[T1]{fontenc} 
\usepackage{amsmath}
\usepackage{verbatim}

\usepackage{amsmath,amssymb, bm}
\usepackage[ruled,vlined,linesnumbered]{algorithm2e}

\SetAlgoVlined
\LinesNumbered
\DontPrintSemicolon
\SetKwInput{KwIn}{Input}
\SetKwInput{KwOut}{Output}
\SetKwInput{KwCons}{Hyperparameters}
\SetKwComment{tcp}{// }{}

\usepackage{csquotes}
\usepackage{multirow}
\usepackage{amsfonts}
\usepackage[dvipsnames]{xcolor}
\definecolor{ACgreen}{HTML}{2E7D32}
\usepackage[export]{adjustbox}
\usepackage{makecell}
\usepackage{tikz}
\usetikzlibrary{shapes.geometric, arrows.meta, positioning}
\tikzstyle{block} = [rectangle, draw, text centered, minimum height=2em, rounded corners=4pt, fill=blue!10]
\tikzstyle{arrow} = [thick, ->, >=Stealth]
\newcolumntype{C}{>{\centering\arraybackslash}X} 
\usepackage[noadjust]{cite}
\usepackage{soul}
\usepackage{xcolor}
\usepackage{url}
\IEEEoverridecommandlockouts

\usepackage[
    colorlinks=true,
    linkcolor=blue,
    citecolor=blue,
    urlcolor=black
]{hyperref}

\begin{document}

\title{Physics-Informed Dynamic State Estimation for Current Transformers Using Graph Neural Networks}

\author{
\IEEEauthorblockN{Michael A. Boateng\IEEEauthorrefmark{1},
Gabriel Gauderman\IEEEauthorrefmark{1}, Nathalie Uwamahoro\IEEEauthorrefmark{2}}

\IEEEauthorblockA{\IEEEauthorrefmark{1} School of Electrical and Computer Engineering, Georgia Institute of Technology
\\Atlanta, Georgia, USA, \{mboateng6, ggauderman3\}@gatech.edu}

\IEEEauthorblockA{\IEEEauthorrefmark{2} Department of Electrical Engineering and Computer Science, Syracuse University
\\Syracuse, New York, USA, nuwamaho@syr.edu}
}

\maketitle
\thispagestyle{plain}
\pagestyle{plain}
\pagenumbering{arabic}
\setcounter{page}{1}

\begin{abstract}
Current transformers are fundamental to power system protection and measurement, yet transient core saturation can severely distort the secondary current and degrade measurement accuracy. Existing dynamic state estimation methods rely mainly on numerical discretisation and iterative solvers, but their initialisation is not informed by the physical dependency structure of the estimation problem, which limits robustness under noisy conditions. This paper presents a physics-informed enhancement for current transformer dynamic state estimation using COMTRADE measurements generated in WinIGS-T. A structured benchmark of four discretisation schemes and three iterative solvers identifies Gauss-Newton with Quadratic discretisation as the strongest baseline. To address the limitation of conventional cold-start initialisation, a graph neural network is constructed from the Jacobian sparsity pattern to generate physics-informed initial state estimates. The proposed warm-start strategy improves estimator conditioning and achieves average gains of 25\% in initialisation distance and 38\% in initial weighted objective value across all tested SNR levels. The results demonstrate that embedding physical structure into the initialisation stage improves the robustness of CT saturation correction and supports more reliable measurement and protection performance in modern power grids.
\end{abstract}

\vspace{0.8em}


\vspace{0.9em}

\section*{Nomenclature}

\subsection{Abbreviations}

\vspace{0.08em}
\begin{tabbing}
    CT \hspace{1.5cm} \= Current transformer. \\
    DSE \> Dynamic state estimation. \\
    MU \> Merging unit. \\
    FE \> Forward Euler. \\
    BE \> Backward Euler. \\
    TR \> Trapezoidal. \\
    QD \> Quadratic. \\
    GN \> Gauss-Newton. \\
    LM \> Levenberg-Marquardt. \\
    SNR \> Signal-to-noise ratio. \\
    GNN \> Graph neural network. \\
    COMTRADE \> Common format for transient data exchange. \\
\end{tabbing}

\vspace{-1.5em}

\subsection{Variables}
\vspace{0.08em}

\begin{tabbing}
    \hspace{2.2cm}\=\hspace{10.5cm}\=\kill
    $n$ \> Number of samples. \\
    $k_{\text{node}}$ \> Node index: $k\in\{1,2,3,4\}$. \\
\end{tabbing}

\vspace{-1.3em}

\begin{tabbing}
\hspace{3.8cm}\=\hspace{10.5cm}\=\kill
    $k_{\text{branch}}$ \> Branch index: $k\in\{1,2,3\}$. \\
    $\lambda_b$ \> Base flux linkage for burden resistor. \\
    $i_b$ \> Base current for burden resistor. \\
    $g_m$ \> Magnetizing conductance for CT. \\
    $L_0$ \> Magnetizing inductance for CT. \\
    $r_1$ \> Primary winding resistance. \\
    $r_2, r_3$ \> Secondary winding resistances. \\
    $L_1, L_2, L_3$ \> CT secondary inductances. \\
    $M_{33}$ \> Mutual inductance for CT model. \\
    $g_{s1}, g_{s2}, g_{s3}$ \> Conductance terms $L_1$--$L_3$. \\
    $R_s$ or $1/g_s$ \> Burden resistor or conductance. \\
    $\lambda(t)$ \> Flux linkage (Weber). \\
    $e(t)$ \> Voltage generated by the flux (V). \\
    $i_p(t)$ \> CT primary current (A). \\
    $i_m(t)$ \> CT magnetizing current (A). \\
    $v_k(t),\;k\in\{1,2,3,4\}$ \> CT secondary-node voltages (V). \\
    $i_{Lk}(t),\;k\in\{1,2,3\}$ \> Currents through $L_1$--$L_3$ (A). \\
\end{tabbing}

\vspace{-1.3em}

\section{Introduction}

\begin{figure}[!t]
    \centering
    \includegraphics[width=1\linewidth]{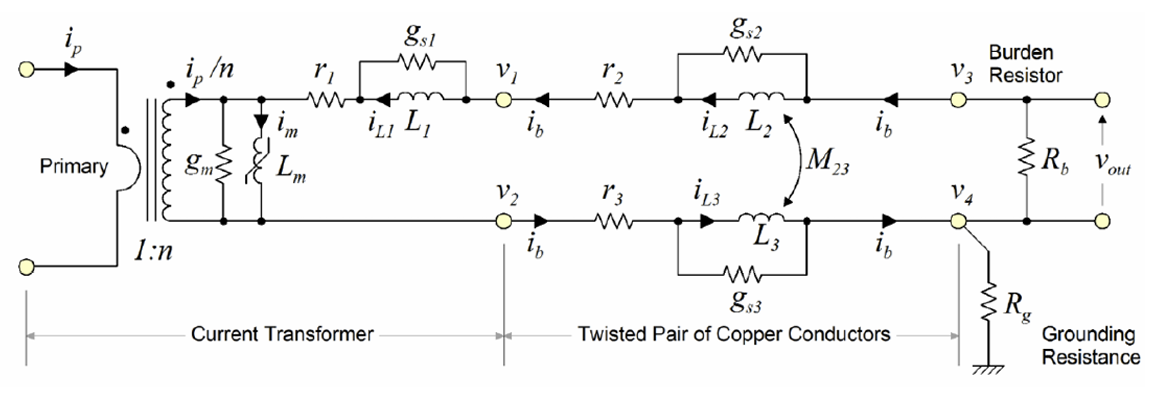}
    \caption{Equivalent circuit of the current-transformer measurement system used in the dynamic state estimation formulation (with burden resistor).}
    \label{fig:CTEquivalentCircuit}
    \vspace{-1em}
\end{figure}

Current transformers~(CTs) are important measurement devices in power system protection and monitoring, yet their accuracy degrades under high fault currents because core saturation distorts the secondary current and can lead to protection maloperation~\cite{hargrave2018beyond}. This concern is especially significant in differential protection, where saturation mismatch across multiple CTs can trigger false operations, including cases caused by inrush currents from adjacent, non-series-connected equipment~\cite{7323856}. Accurate CT error estimation is therefore critical for reliable grid operation. To address this need, the present work benchmarks DSE methods for CT error compensation and proposes a physics-informed DSE framework that leverages graph neural networks. Fig.~\ref{fig:CTEquivalentCircuit} shows the equivalent circuit of the current transformer.

CT error estimation has been studied extensively through physics-based dynamic modeling. Nonlinear transient CT models were developed to represent CT behavior during faults by incorporating burden impedance, magnetic-core saturation, and other time-varying effects~\cite{naidu1986dynamic,tziouvaras2000mathematical}. These studies established that CT dynamic errors arise from intrinsic physical nonlinearities and therefore cannot be represented adequately by static correction factors. Their main contribution lies in providing physically grounded models for relay studies, transient simulation, and error characterization. However, these approaches remain primarily forward models whose performance depends on accurate parameterization and whose formulation does not directly provide a robust inverse framework for recovering the true current or CT error state from noisy measurements in real time.

To improve practical CT accuracy, subsequent studies introduced digital compensation, calibration, and distortion-aware correction techniques. Frequency-response-based compensation, online waveform reconstruction, harmonic-distortion correction, and tensor-linearization methods demonstrated improved CT performance under harmonics, moderate nonlinearities, and distorted operating conditions~\cite{gallo2010real,haghjoo2015compensation,ballal2018novel,collin2019compensation}. Experimental investigations further showed that distorted network conditions, frequency-dependent magnetic-branch behavior, and realistic disturbances can substantially affect CT accuracy, thereby exposing the limitations of standard steady-state testing~\cite{mingotti2020inductive,stano2020wideband,laurano2021simple,stano2022understanding,krzysztof2025review}. Despite these advances, most compensation and calibration methods remain limited by model-identification quality, sensitivity to operating-condition mismatch, and the absence of a dedicated mechanism for event-by-event inversion under noisy transient conditions.

More recent studies introduced machine-learning-based methods for CT error prediction, waveform restoration, and saturation detection. Deep learning architectures that combine convolutional, attention, and recurrent components formulated CT error estimation as a temporal learning problem, while artificial neural networks and LSTM-based models were used to restore distorted secondary currents and detect saturation under noisy conditions~\cite{li2024prediction,odinaev2024restoration,danquah2025machine,ren2025estimation}. Signal-processing-based detectors, including cross-correlation methods, also improved the identification of saturation intervals without requiring CT-specific parameters or additional hardware~\cite{elgamasy2025current}. These approaches improved prediction, detection, and restoration, particularly when traditional threshold-based or harmonic-based techniques become sensitive to noise. However, most existing learning-based methods still operate at the waveform or device level, depend strongly on the representativeness of the training data, and do not explicitly enforce CT electromagnetic constraints or incorporate the electrical network structure associated with a broader estimation problem.

In parallel, the broader power-systems literature has emphasized robust dynamic state estimation for time-varying and noisy operating conditions. Dynamic state estimation methods and robust filter-based formulations demonstrated improved performance under non-Gaussian noise, impulsive disturbances, and outliers~\cite{zhao2019power,chen2024new}. Despite these contributions, the problem remains dispersed across separate tasks of CT characterization, correction, detection, and general estimation. A clear need therefore remains for a hybrid CT-error-estimation framework that preserves the physical consistency of a nonlinear estimator while incorporating topology-aware prior information from graph-based learning. Accordingly, the contributions of this work are as follows:

\begin{itemize}
    \item A comprehensive benchmark of integration methods (Forward Euler, Trapezoidal, Quadratic) and iterative solvers (Gauss-Newton, Levenberg-Marquardt, Hybrid GN→LM) for dynamic state estimation in CT saturation correction, identifying optimal solver-integration pairings for speed, accuracy, and robustness.
    
    \item A physics-informed graph neural network warm-start strategy that improves convergence initialization quality while maintaining comparable post-convergence accuracy to traditional cold-start approaches.
    
\end{itemize}

\section{Methodology}
\label{sec:methodology}

This section presents the complete methodology for integrating a graph neural network (GNN) initializer with Gauss--Newton (GN) optimization in CT dynamic state estimation. We begin with the integration and solver foundations, then introduce the physics-informed GNN architecture and training.

\subsection{Integration Methods for Discrete-Time DSE}
\label{sec:integration_methods}

The CT DSE problem requires discretizing continuous-time state dynamics as outlined in Model~1 (Eqs.~\eqref{eq:CTDSE-int}--\eqref{eq:CTDSE-qd}). This work evaluates four integration schemes across SNR levels to identify computational and accuracy trade-offs. 

\subsubsection{Forward Euler}
Forward Euler (Eq.~\eqref{eq:CTDSE-fe}) is the simplest and fastest integration method, achieving $\mathcal{O}(h)$ accuracy per step with only a single function evaluation per iteration. However, explicit schemes can suffer stability issues under stiff dynamics or large step sizes.

\subsubsection{Backward Euler}
The implicit backward Euler scheme (Eq.~\eqref{eq:CTDSE-be}) offers superior stability for stiff problems and is unconditionally stable but requires solving an implicit equation at each step, increasing computational cost.

\subsubsection{Trapezoidal Rule}
The trapezoidal integration method (Eq.~\eqref{eq:CTDSE-tr}) improves accuracy by averaging endpoint derivatives, achieving $\mathcal{O}(h^2)$ local truncation error and providing improved stability compared to forward Euler, making it a standard choice in many industrial DSE applications.

\subsubsection{Quadratic (Simpson Variant)}
The quadratic discretization (Eq.~\eqref{eq:CTDSE-qd}) extends accuracy by using a second-order polynomial approximation of the state trajectory, providing $\mathcal{O}(h^3)$ local accuracy and is particularly effective when state derivatives are available or can be efficiently computed.

\subsection{Integration Solvers for Discrete-Time DSE}
\label{sec:iterative_solvers}

After discretizing the state dynamics using one of the above schemes, the resulting nonlinear estimation problem is solved iteratively. This work compares three solver variants specified in Model~1 (Eqs.~\eqref{eq:CTDSE-gn}--\eqref{eq:CTDSE-switch}).

\subsubsection{Gauss--Newton}
The standard Gauss--Newton method (Eq.~\eqref{eq:CTDSE-gn}) solves the weighted least-squares problem at each iteration. Gauss--Newton converges rapidly at high SNR but can diverge when the initial guess is far from the solution or when measurement noise is substantial.

\subsubsection{Levenberg--Marquardt}
The Levenberg--Marquardt method (Eq.~\eqref{eq:CTDSE-lm}) adds adaptive damping to improve robustness. LM is significantly more robust at low SNR but typically requires $\mathrm{20}$--$\mathrm{30}\%$ more computation than GN due to the damping search and matrix re-factorizations.

\subsubsection{Hybrid GN$\rightarrow$LM Strategy}
To balance speed and robustness, a hybrid approach is employed (Eq.~\eqref{eq:CTDSE-switch}): start with Gauss--Newton and automatically switch to Levenberg--Marquardt if divergence is detected (e.g., residual increases between iterations).

\subsection{Physics-Informed Graph Neural Network Architecture}
\label{sec:gnn_architecture}

To improve initialization quality for the iterative solvers, a graph neural network (GNN) is trained to map noisy measurements to informed state estimates (Eq.~\eqref{eq:CTDSE-init}). The CT state is represented as a directed graph with edges from the Jacobian sparsity pattern. Message-passing layers propagate information across physical dependencies to learn state couplings. The GNN is trained on $\mathrm{2000}$ samples using reference states from high-SNR ($+\mathrm{20}$ dB) Gauss--Newton convergence. The loss function combines MSE and $L_{\mathrm{2}}$ regularization:

{\small
\begin{equation}
\mathcal{L} = \frac{1}{|\mathcal{B}|} \sum_{k \in \mathcal{B}} \left\| \hat{\mathbf{x}}_k^{\mathrm{GNN}} - \mathbf{x}_k^{\mathrm{ref}} \right\|_2^2 + w_{\mathrm{reg}} \|\boldsymbol{\theta}\|_2^2
\label{eq:gnn_loss}
\end{equation}
}

\noindent with $w_{\mathrm{reg}} = 0.001$, training uses Adam (learning rate 0.01), batch size 32, and 20 epochs.

\subsubsection{Warm-Start Gauss--Newton with GNN Initialization}

Algorithm~\ref{alg:gnn_warm_start} specifies the complete warm-start procedure. Instead of initializing with $\mathbf{x}^{(0)} = \mathbf{0}$, the Gauss--Newton solver receives the GNN output as specified in Eq.~\eqref{eq:CTDSE-init}: $\mathbf{x}^{(0)} = \Phi_{\mathrm{GNN}}(\mathcal{G}_k,\mathbf{z}_k,\hat{\mathbf{x}}_{k|k-1})$. Subsequent GN iterations proceed with quadratic integration until convergence.

\begin{figure}[!t]
\centering
\begin{minipage}{\linewidth}
\noindent\rule{\linewidth}{0.6pt}

\vspace{-0.2em}
\noindent \textbf{Model 1:} Physics-Informed CT Dynamic State Estimation with Switchable Integration and Iterative Solvers.

\vspace{-0.6em}
\noindent\rule{\linewidth}{0.3pt}

\vspace{0.15em}

\begin{small}
\noindent\textbf{States / variables: } $\mathbf{x}_k$ (dynamic state), $\Delta \mathbf{x}_k$ (state correction), $\mathbf{r}_k$ (residual), $\mathbf{J}_k$ (Jacobian); \textbf{Inputs: } $\mathbf{z}_k$ (measurement vector), $\mathbf{u}_k$ (known inputs); \textbf{Choices: } $m_k \in \{\mathrm{FE},\mathrm{BE},\mathrm{TR},\mathrm{QD}\}$ (integration rule), $s_k \in \{\mathrm{GN},\mathrm{LM},\mathrm{GN}\!\rightarrow\!\mathrm{LM}\}$ (iterative solver); \textbf{Parameters: } $h=\Delta t$, $\lambda_k$ (LM damping), $\ell_{\mathrm{sw}}$ (GN-to-LM switch iteration).
\end{small}

\vspace{-1.0em}

\begin{small}
\begin{subequations}
\label{eq:CTDSE}
\begin{align}
    & \hspace{-0.8em}\textbf{Measurement Model:} \nonumber \\
    & \mathbf{z}_k = \mathbf{h}(\mathbf{x}_k,\mathbf{x}_{k-1}) + \bm{\varepsilon}_k,
    \qquad
    \bm{\varepsilon}_k \sim \mathcal{N}(\mathbf{0},\mathbf{R}_k)
    \label{eq:CTDSE-meas} \\[0.25em]
    & \hspace{-0.8em}\textbf{State Propagation with Selectable Integrator:} \nonumber \\
    & \hat{\mathbf{x}}_{k|k-1}
    =
    \mathcal{I}_{m_k}\!\left(\hat{\mathbf{x}}_{k-1},\mathbf{u}_{k-1},h\right),
    \qquad
    m_k \in \{\mathrm{FE},\mathrm{BE},\mathrm{TR},\mathrm{QD}\}
    \label{eq:CTDSE-int} \\
    & \mathcal{I}_{\mathrm{FE}}:\;
    \mathbf{x}_{k}
    =
    \mathbf{x}_{k-1}
    + h\,\mathbf{f}(\mathbf{x}_{k-1},\mathbf{u}_{k-1})
    \label{eq:CTDSE-fe} \\
    & \mathcal{I}_{\mathrm{BE}}:\;
    \mathbf{x}_{k}
    =
    \mathbf{x}_{k-1}
    + h\,\mathbf{f}(\mathbf{x}_{k},\mathbf{u}_{k})
    \label{eq:CTDSE-be} \\
    & \mathcal{I}_{\mathrm{TR}}:\;
    \mathbf{x}_{k}
    =
    \mathbf{x}_{k-1}
    + \frac{h}{2}\!\left[
    \mathbf{f}(\mathbf{x}_{k-1},\mathbf{u}_{k-1})
    +
    \mathbf{f}(\mathbf{x}_{k},\mathbf{u}_{k})
    \right]
    \label{eq:CTDSE-tr} \\
    & \mathcal{I}_{\mathrm{QD}}:\;
    \mathbf{x}_{k}
    =
    \mathbf{x}_{k-1}
    + h\,\mathbf{f}(\mathbf{x}_{k-1},\mathbf{u}_{k-1})
    + \frac{h^2}{2}\,\dot{\mathbf{f}}(\mathbf{x}_{k-1},\mathbf{u}_{k-1})
    \label{eq:CTDSE-qd} \\[0.25em]
    & \hspace{-0.8em}\textbf{Initialization / Warm Start:} \nonumber \\
    & \mathbf{x}_k^{(0)}
    =
    \begin{cases}
        \hat{\mathbf{x}}_{k|k-1}, & \text{cold start} \\[0.2em]
        \Phi_{\mathrm{GNN}}(\mathcal{G}_k,\mathbf{z}_k,\hat{\mathbf{x}}_{k|k-1}),
        & \text{GNN warm start}
    \end{cases}
    \label{eq:CTDSE-init} \\[0.25em]
    & \hspace{-0.8em}\textbf{Weighted Residual and Objective:} \nonumber \\
    & \mathbf{r}_k^{(\ell)}
    =
    \mathbf{z}_k - \mathbf{h}\!\left(\mathbf{x}_k^{(\ell)},\mathbf{x}_{k-1}\right),
    \qquad
    \mathbf{J}_k^{(\ell)}
    =
    \frac{\partial \mathbf{h}}{\partial \mathbf{x}}
    \Bigg|_{\mathbf{x}=\mathbf{x}_k^{(\ell)}}
    \label{eq:CTDSE-res} \\
    & J_k\!\left(\mathbf{x}_k^{(\ell)}\right)
    =
    \frac{1}{2}\,
    \mathbf{r}_k^{(\ell)\mathsf{T}}
    \mathbf{W}_k
    \mathbf{r}_k^{(\ell)},
    \qquad
    \mathbf{W}_k = \mathbf{R}_k^{-1}
    \label{eq:CTDSE-obj} \\[0.25em]
    & \hspace{-0.8em}\textbf{Iterative Update Rules:} \nonumber \\
    & \mathcal{U}_{\mathrm{GN}}
    =
    \left(\mathbf{J}_k^{\mathsf T}\mathbf{W}_k\mathbf{J}_k\right)^{-1}
    \mathbf{J}_k^{\mathsf T}\mathbf{W}_k\mathbf{r}_k
    \label{eq:CTDSE-gn} \\
    & \mathcal{U}_{\mathrm{LM}}
    =
    \left(\mathbf{J}_k^{\mathsf T}\mathbf{W}_k\mathbf{J}_k + \lambda_k\mathbf{I}\right)^{-1}
    \mathbf{J}_k^{\mathsf T}\mathbf{W}_k\mathbf{r}_k
    \label{eq:CTDSE-lm} \\
    & \Delta\mathbf{x}_k^{(\ell)}
    =
    \begin{cases}
        \mathcal{U}_{\mathrm{GN}}, & s_k=\mathrm{GN} \\[0.2em]
        \mathcal{U}_{\mathrm{LM}}, & s_k=\mathrm{LM} \\[0.2em]
        \mathcal{U}_{\mathrm{GN}}, & s_k=\mathrm{GN}\!\rightarrow\!\mathrm{LM},\ \ell \le \ell_{\mathrm{sw}} \\[0.2em]
        \mathcal{U}_{\mathrm{LM}}, & s_k=\mathrm{GN}\!\rightarrow\!\mathrm{LM},\ \ell > \ell_{\mathrm{sw}}
    \end{cases}
    \label{eq:CTDSE-switch} \\
    & \mathbf{x}_k^{(\ell+1)}
    =
    \mathbf{x}_k^{(\ell)} + \Delta\mathbf{x}_k^{(\ell)}
    \label{eq:CTDSE-stateupdate}
\end{align}
\end{subequations}
\end{small}

\vspace{-0.45em}

\begin{small}
\textbf{Convergence: } $\|\Delta \mathbf{x}_k^{(\ell)}\|_\infty < \epsilon_x$ or $\|\mathbf{r}_k^{(\ell)}\|_\infty < \epsilon_r$, with final estimate $\hat{\mathbf{x}}_k = \mathbf{x}_k^{(\ell^\star)}$.
\end{small}

\vspace{-0.4em}
\noindent\rule{\linewidth}{0.3pt}
\end{minipage}

\label{fig:model_ctdse_switching}
\vspace{-1.2em}
\end{figure}

\section{Numerical Experiments and Discussions}
\label{sec:num_exp_discuss}

The integration and solver variants described in Section~\ref{sec:methodology} were evaluated through systematic benchmarking using supervised learning on COMTRADE measurements. This section presents the computational setup, dataset preparation, training configuration, and  result analysis stratified by SNR level.

\subsection{Experimental Setup and Dataset Preparation}
\label{sec:exp_setup}

Experiments were conducted on a laptop powered by an 11th Gen Intel Core i5-1135G7 processor clocked at 2.40~GHz, with 12~GB of RAM and integrated Intel Iris Xe Graphics. This configuration is representative of portable development and prototyping environments where practitioners validate algorithms before real-time deployment. The CT circuit data were generated using WinIGS-T (Integrated Grounding System Analysis program for Windows)~\cite{advanced-power-concepts-2024}, producing high-fidelity COMTRADE recordings reflecting field conditions during saturation. These measurements, representing burden voltage, allow traditional reconstruction of the primary current using known burden resistance and turns ratio.

\begin{figure}[!t]
    \centering
    \includegraphics[width=1\linewidth]{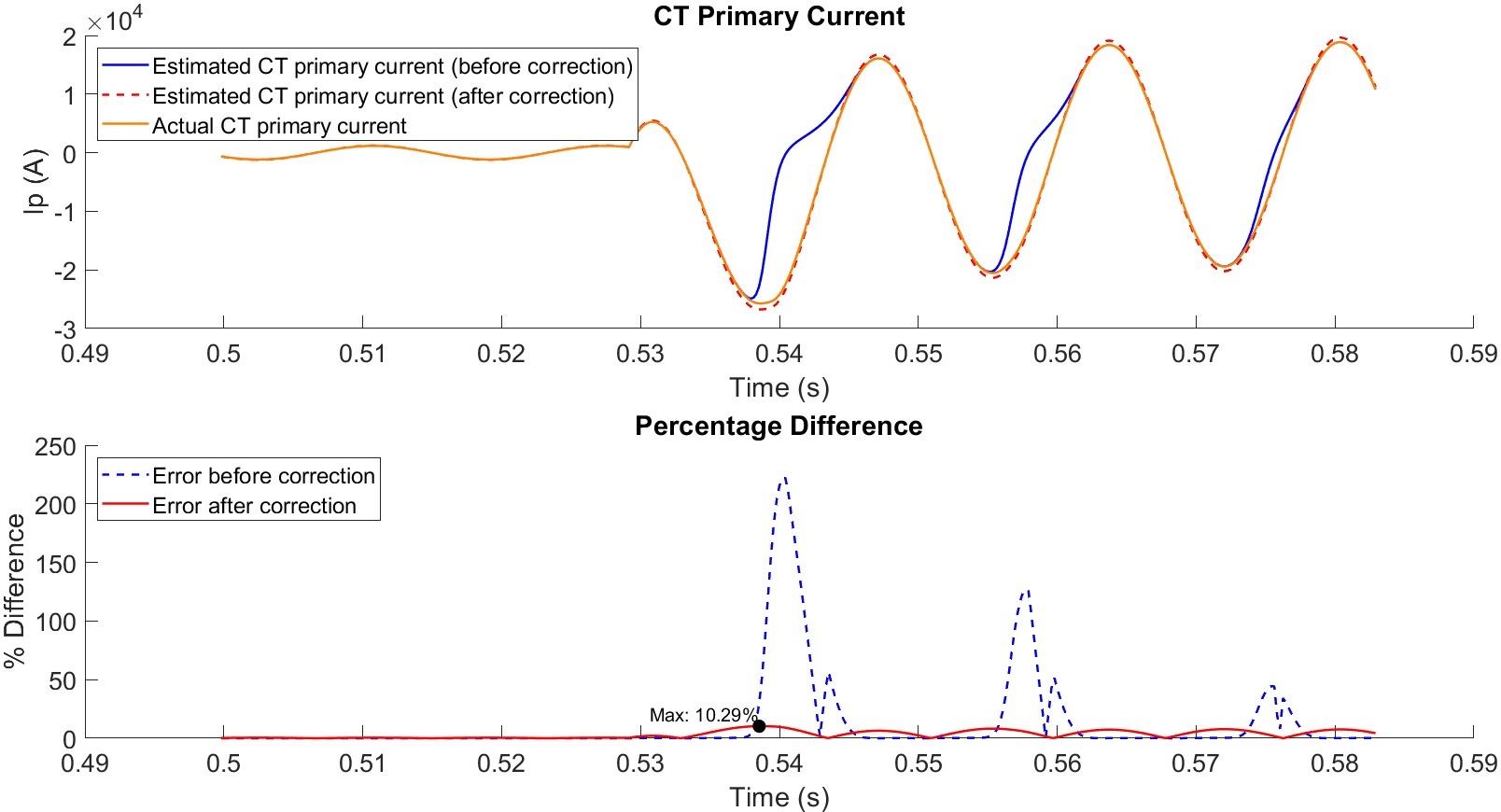}
    \caption{Actual and estimated CT primary current waveforms before and after correction, with the associated percentage error curves (max error shown).}
    \label{fig:Ev1Error}
    \vspace{-1em}
\end{figure}

Figure~\ref{fig:Ev1Error} demonstrates the severity of CT saturation: the reconstructed current (blue curve, ``before correction'') deviates substantially from the actual current (yellow curve), yielding maximum errors exceeding $\mathbf{200}\%$ during saturation events (bottom half). This motivates the need for robust estimation algorithms that can recover the true primary current despite measurement degradation.

\begin{algorithm}[!t]
\small
\caption{GNN Warm-Start Execution for CT Dynamic State Estimation}
\label{alg:gnn_warm_start}

\KwIn{%
  Measurement samples $\{\mathbf{z}_k\}_{k=1}^{K}$ and reference states $\{\mathbf{x}_k^{\mathrm{ref}}\}_{k=1}^{K}$\\
  Adjacency list $\mathcal{A}$ for the 15-state CT graph\\
  SNR set $\mathcal{S}_{\mathrm{snr}}=\{20,15,10,5,0,-5,-10\}$ dB
}
\KwOut{%
  Trained GNN initializer $\Phi_{\mathrm{GNN},\theta^\star}$\\
  Normalization statistics $(\boldsymbol{\mu}_z,\boldsymbol{\sigma}_z,\boldsymbol{\mu}_x,\boldsymbol{\sigma}_x)$\\
  Benchmark table stratified by SNR
}
\KwCons{%
  Hidden size $d_h=32$; GNN layers $L=2$; batch size $B=32$;\\
  Learning rate $\eta=10^{-2}$; epochs $N_e=20$;\\
  GN tolerance $\epsilon=10^{-6}$; max iterations $I_{\max}=60$
}

\tcp{(A) Normalize training data}
Compute per-channel statistics $\boldsymbol{\mu}_z,\boldsymbol{\sigma}_z$ from $\{\mathbf{z}_k\}$ and per-state statistics $\boldsymbol{\mu}_x,\boldsymbol{\sigma}_x$ from $\{\mathbf{x}_k^{\mathrm{ref}}\}$\;

$\tilde{\mathbf{z}}_k \gets (\mathbf{z}_k-\boldsymbol{\mu}_z)\oslash(\boldsymbol{\sigma}_z+10^{-8})$,\quad
$\tilde{\mathbf{x}}_k \gets (\mathbf{x}_k^{\mathrm{ref}}-\boldsymbol{\mu}_x)\oslash(\boldsymbol{\sigma}_x+10^{-8})$\;

\tcp{(B) Train GNN initializer}
Initialize $\Phi_{\mathrm{GNN},\theta}$ with embedding $(1\rightarrow d_h)$, $L$ message-passing layers on $\mathcal{A}$, and readout $(d_h\rightarrow 1)$ per node\;

\For{$e=1,\dots,N_e$}{
  Shuffle samples and form mini-batches of size $B$\;
  \ForEach{mini-batch $(\tilde{\mathbf{Z}},\tilde{\mathbf{X}})$}{
    $\hat{\mathbf{X}} \gets \Phi_{\mathrm{GNN},\theta}(\tilde{\mathbf{Z}})$\;
    $\mathcal{L} \gets \mathrm{MSE}(\hat{\mathbf{X}},\tilde{\mathbf{X}})$\;
    $\theta \gets \theta - \eta \nabla_{\theta}\mathcal{L}$\;
  }
}
$\theta^\star \gets \theta$\;

\tcp{(C) Evaluate warm-start versus cold-start GN across SNR}
\ForEach{$\mathrm{snr}\in\mathcal{S}_{\mathrm{snr}}$}{
  Build noisy measurements $\mathbf{z}_k^{\mathrm{noisy}} = \mathbf{z}_k + \boldsymbol{\eta}_k$ with $\boldsymbol{\eta}_k \sim \mathcal{N}(0,\sigma_{\mathrm{snr}}^2\mathbf{I})$\;
  Initialize accumulators for iterations, objective values, convergence, and initialization quality\;

  \For{$k=1,\dots,N_{\mathrm{test}}$}{
    \tcp{Cold start}
    $\mathbf{x}_{k,\mathrm{init}}^{\mathrm{cold}} \gets \mathbf{0}$\;

    \tcp{GNN warm start}
    $\tilde{\mathbf{z}}_k \gets (\mathbf{z}_k^{\mathrm{noisy}}-\boldsymbol{\mu}_z)\oslash(\boldsymbol{\sigma}_z+10^{-8})$\;
    $\tilde{\mathbf{x}}_{k,\mathrm{init}}^{\mathrm{gnn}} \gets \Phi_{\mathrm{GNN},\theta^\star}(\tilde{\mathbf{z}}_k)$\;
    $\mathbf{x}_{k,\mathrm{init}}^{\mathrm{gnn}} \gets
    \tilde{\mathbf{x}}_{k,\mathrm{init}}^{\mathrm{gnn}}\odot\boldsymbol{\sigma}_x + \boldsymbol{\mu}_x$\;

    \tcp{Run GN solver twice using identical settings}
    \mbox{$\left(\hat{\mathbf{x}}_k^{\mathrm{cold}}, N_k^{\mathrm{cold}}, \mathcal{J}_k^{\mathrm{cold}}\right)
    \gets \mathrm{GN}\!\left(\mathbf{z}_k^{\mathrm{noisy}},\mathbf{x}_{k,\mathrm{init}}^{\mathrm{cold}},\epsilon,I_{\max}\right)$}\;

    \mbox{$\left(\hat{\mathbf{x}}_k^{\mathrm{gnn}}, N_k^{\mathrm{gnn}}, \mathcal{J}_k^{\mathrm{gnn}}\right)
    \gets \mathrm{GN}\!\left(\mathbf{z}_k^{\mathrm{noisy}},\mathbf{x}_{k,\mathrm{init}}^{\mathrm{gnn}},\epsilon,I_{\max}\right)$}\;

    Record metrics:
    $N_k^{\mathrm{cold}}$, $N_k^{\mathrm{gnn}}$,
    $\mathcal{J}_k^{\mathrm{cold}}$, $\mathcal{J}_k^{\mathrm{gnn}}$,\;
    \mbox{$\|\mathbf{x}_{k,\mathrm{init}}^{\mathrm{cold}}-\mathbf{x}_k^{\mathrm{ref}}\|_2$,
    $\|\mathbf{x}_{k,\mathrm{init}}^{\mathrm{gnn}}-\mathbf{x}_k^{\mathrm{ref}}\|_2$}\;
  }

  Aggregate by SNR: mean iterations, iteration gain (\%), initialization-distance gain (\%), objective gain (\%), and convergence rate\;
}

\KwRet{$\Phi_{\mathrm{GNN},\theta^\star}$, $(\boldsymbol{\mu}_z,\boldsymbol{\sigma}_z,\boldsymbol{\mu}_x,\boldsymbol{\sigma}_x)$, and the benchmark table}
\end{algorithm}

\subsubsection{Data Preparation and Noise Model}

The dataset consists of COMTRADE measurements, partitioned into $\mathrm{2{,}000}$ samples for GNN training and $\mathrm{1{,}440}$ samples (split $\mathrm{60}/\mathrm{40}$ into test and validation sets) for benchmarking integration and solver methods. Measurement noise was systematically injected across SNR levels $\mathcal{S}_{\mathrm{snr}}=\{-\mathrm{10},-\mathrm{5},\mathrm{0},\mathrm{5},\mathrm{10},\mathrm{15},\mathrm{20}\}\,\mathrm{dB}$ using Gaussian additive noise $\boldsymbol{\epsilon}\sim\mathcal{N}(\mathbf{0},\sigma^2)$, where $\sigma$ was selected to match each SNR level. Reference states for GNN training were obtained by running Gauss--Newton to convergence on high-SNR ($+\mathrm{20}\,\mathrm{dB}$) measurements, thereby providing reference state estimates that the GNN was trained to approximate.

\subsubsection{GNN Training Configuration}

The GNN was trained with hidden dimension $d_h=\mathrm{32}$ for lightweight embedding suitable for embedded systems and $L=\mathrm{2}$ message-passing layers to balance expressiveness and computational cost. Training employed the Adam optimizer with learning rate $\eta=\mathrm{0.01}$ and exponential decay, using batch size $B=\mathrm{32}$ over $\mathrm{20}$ epochs with early stopping based on validation loss. Regularization was applied with weight $w_{\mathrm{reg}}=\mathrm{0.001}$ to reduce overfitting. All input data were normalized to zero mean and unit variance using training-set statistics, and no layer normalization was applied.

\subsubsection{Evaluation Metrics}

Algorithm performance was evaluated using four primary metrics: average solving time (ms), defined as the wall-clock time per sample; convergence rate (\%), defined as the percentage of samples converging within $\mathrm{15}$ iterations; average residual, defined as the mean weighted residual norm at convergence; and average iterations, defined as the mean iteration count to convergence. For GNN warm-start analysis, the relative gains are defined as:

\begin{subequations}\label{eq:gnn_gain_metrics}
\begin{align}
\Delta_{\mathrm{dist}}
&= \mathrm{100}\!\left(1-\frac{\|\mathbf{x}_0^{\mathrm{GNN}}-\mathbf{x}^{\mathrm{ref}}\|_2}{\|\mathbf{0}-\mathbf{x}^{\mathrm{ref}}\|_2}\right),
\label{eq:gnn_gain_dist}\\
\Delta_{J}
&= \mathrm{100}\!\left(1-\frac{J_0^{\mathrm{GNN}}}{J_0^{\mathrm{cold}}}\right),
\label{eq:gnn_gain_obj}\\
\Delta_{\mathrm{iters}}
&= \mathrm{100}\!\left(1-\frac{N_{\mathrm{GNN}}}{N_{\mathrm{cold}}}\right),
\label{eq:gnn_gain_iters}
\end{align}
\end{subequations}

where \eqref{eq:gnn_gain_dist} denotes initialization-distance gain, \eqref{eq:gnn_gain_obj} denotes initialization-objective gain, and \eqref{eq:gnn_gain_iters} denotes iter. reduction.

\subsection{Integration Methods Assessment}
\label{sec:results_integration}

\begin{figure*}[!t]
    \centering
    \includegraphics[width=1\textwidth]{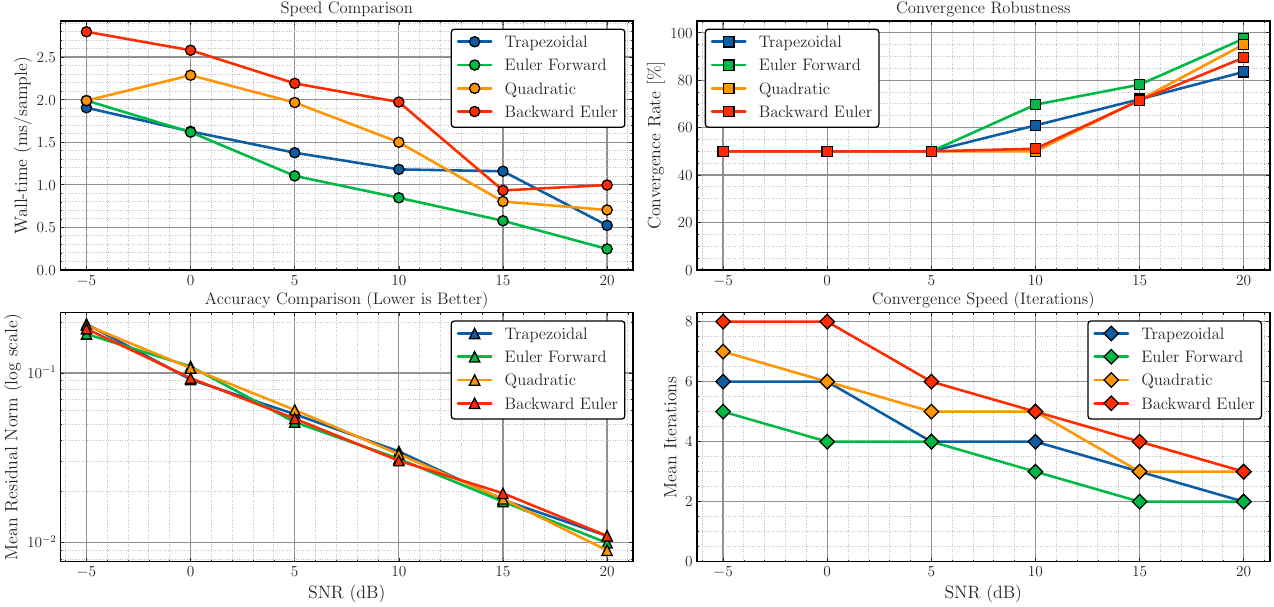}
    \vspace{-2em}
    \caption{\text{Integration-method benchmark under varying $\mathrm{SNR}$ (GN solver).}
    Four panels report: wall-time (top-left), convergence rate (top-right), residual norm in log scale (bottom-left), and mean iterations (bottom-right), each versus $\mathrm{SNR}$.
    Curves compare
    \textbf{\textcolor{blue}{Trapezoidal}},
    \textbf{\textcolor{ACgreen}{Euler Forward}},
    \textbf{\textcolor{orange}{Quadratic}}, and
    \textbf{\textcolor{red}{Backward Euler}}.
    The figure highlights the practical trade-off between speed, robustness, and accuracy as noise increases.}
    \label{fig:integration_methods}
    \vspace{-0.8em}
\end{figure*}

\textit{1) Computational Efficiency and Accuracy Trade-off:}
Figure~\ref{fig:integration_methods} presents a four-panel benchmark of integration schemes (Forward Euler, Trapezoidal, Quadratic, Backward Euler) under varying $\mathrm{SNR}$, isolating integration effects by using a fixed Gauss--Newton solver. Panel \textit{(a)} shows wall-clock execution time per sample: Forward Euler is fastest at $\sim 1\,\mathrm{ms/sample}$, while Backward Euler is slowest at $\sim 2\,\mathrm{ms/sample}$ because of implicit equation solving. This $48\%$--$79\%$ overhead for higher-order schemes must be weighed against modest accuracy gains. Panel \textit{(b)} shows convergence-rate degradation at low $\mathrm{SNR}$ for all methods, with Trapezoidal maintaining $\sim 61\%$ convergence down to $\mathrm{SNR}=-5\,\mathrm{dB}$. Forward Euler is slightly more robust ($66\%$) but less accurate overall. Panel \textit{(c)} reports the residual norm in log scale, where all methods reach similar final accuracy; Quadratic gives only marginal improvement at $\mathrm{SNR}\geq 0\,\mathrm{dB}$, which is usually insufficient to justify its extra cost. Panel \textit{(d)} shows mean iterations to convergence, ranging from $3.3$ (Forward Euler) to $5.7$ (Backward Euler), indicating that integration order does not strongly affect convergence speed once $\mathrm{SNR}$ is fixed.

Trapezoidal emerges as the preferred general-purpose choice, balancing about $22\%$ extra cost with robust $\sim 61\%$ convergence across noise levels. Higher-order schemes are better suited to offline post-processing where accuracy matters more than runtime.

\subsection{Iterative Solver Robustness Analysis}
\label{sec:results_solvers}

\begin{figure*}[!t]
    \centering
    \includegraphics[width=1\textwidth]{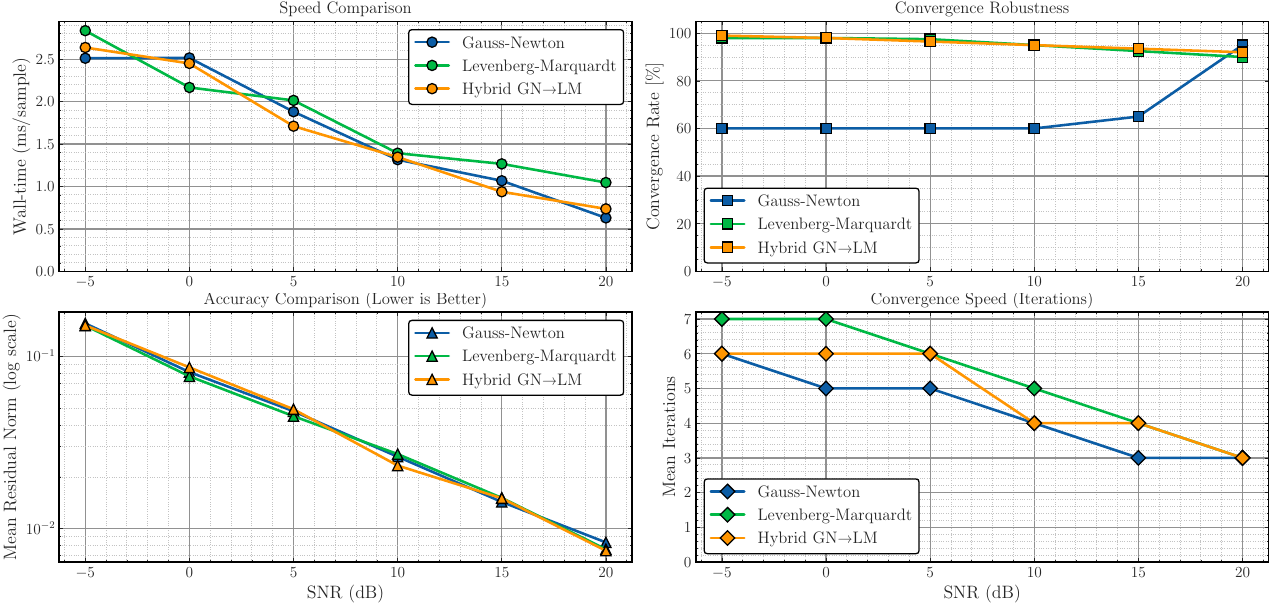}
    \vspace{-2em}
    \caption{\text{Iterative-solver benchmark across SNR levels.}
    Panels show wall-time (top-left), convergence rate (top-right), residual norm on a log scale (bottom-left), and mean iterations (bottom-right).
    Results are shown for
    \textbf{\textcolor{blue}{Gauss--Newton}},
    \textbf{\textcolor{ACgreen}{Levenberg--Marquardt}}, and
    \textbf{\textcolor{orange}{Hybrid GN$\rightarrow$LM}}.
    The comparison emphasizes solver robustness at low SNR and computational efficiency at high SNR.}
    \label{fig:iterative_solvers}
    \vspace{-1.2em}
\end{figure*}

\textit{1) Convergence Robustness Across Noise Regimes:}
Figure~\ref{fig:iterative_solvers} compares Gauss--Newton, Levenberg--Marquardt, and Hybrid GN$\rightarrow$LM using Quadratic integration. Panel \textit{(a)} shows execution time: Gauss--Newton is fastest at $\sim 1.6\,\mathrm{ms/sample}$, while Levenberg--Marquardt incurs about $8\%$ overhead ($\sim 1.79\,\mathrm{ms}$) due to damping adaptation. Panel \textit{(b)} reveals the main robustness gap: Gauss--Newton achieves only $67\%$ convergence at low $\mathrm{SNR}$, indicating frequent divergence under strong noise. In contrast, both Levenberg--Marquardt and Hybrid GN$\rightarrow$LM maintain about $95\%$ convergence across all $\mathrm{SNR}$ levels, showing that adaptive damping is essential in noisy conditions. Panel \textit{(c)} shows final residual accuracy in log scale, where LM and Hybrid produce slightly lower residuals ($0.054$--$0.056$) than GN ($0.056$), indicating marginally better final-state accuracy. Panel \textit{(d)} reports iteration counts: Gauss--Newton requires $4.3$ iterations on average, while Levenberg--Marquardt requires $5.3$, reflecting the added damping-search effort.

\textit{2) Selection Guidance for Deployment:}
The Hybrid GN$\rightarrow$LM strategy provides the best overall balance, matching LM-level robustness ($95.7\%$ convergence) while reducing computational cost ($4.8$ iterations, $1.64\,\mathrm{ms}$). It is therefore recommended for variable-$\mathrm{SNR}$ operating conditions where robustness cannot be sacrificed.

\subsection{GNN Warm-Start Initialization Impact}
\label{sec:results_gnn}

\begin{figure*}[!t]
    \centering
    \includegraphics[width=1\textwidth]{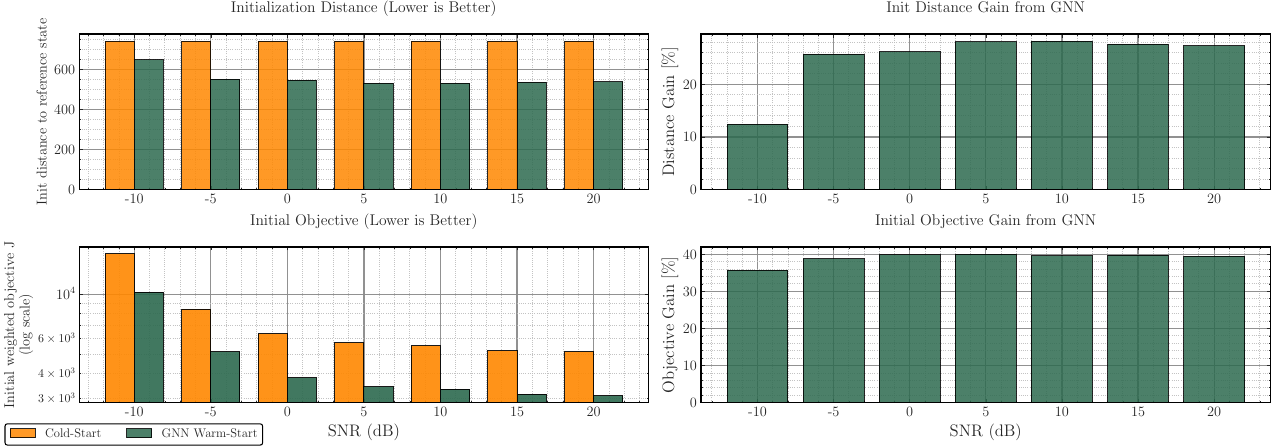}
    \caption{\text{GNN warm-start improves initialization quality for GN.}
    Top-left: initialization distance to reference state (lower is better);
    top-right: distance gain (\%);
    bottom-left: initial weighted objective $J$ (log scale, lower is better);
    bottom-right: objective gain (\%).
    Bar groups compare
    \textbf{\textcolor{orange}{Cold-Start}} and
    \textbf{\textcolor{ACgreen}{GNN Warm-Start}}.
    The figure shows that GNN primarily strengthens the initialization stage, especially under noisy conditions.}
    \label{fig:gnn_init_advantage}
    \vspace{-0.8em}
\end{figure*}

\textit{1) Initialization Quality Improvement:}
Figure~\ref{fig:gnn_init_advantage} isolates the GNN initialization effect by comparing cold-start ($\mathbf{x}_0=\mathbf{0}$) and GNN warm-start ($\mathbf{x}_0=\Phi_{\mathrm{GNN}}(\mathbf{z})$) using the same Gauss--Newton solver with Quadratic integration. Panel \textit{(a)} shows that the GNN initialization is consistently closer to the reference state, typically by $200$--$250$ units across $\mathrm{SNR}$ levels, with the best result near $\mathrm{SNR}=10\,\mathrm{dB}$. Panel \textit{(b)} quantifies this gain: the average initialization-distance reduction is $\mathbf{25.04}\%$, reaching $\mathbf{28.13}\%$ at moderate $\mathrm{SNR}$ and dropping to $\mathbf{12.36}\%$ at the lowest $\mathrm{SNR}$. This indicates that the GNN provides a stable regularizing prior even under strong noise. Panel \textit{(c)} reports the initial weighted objective $\mathcal{J}$ in log scale. The GNN begins with a $3$--$10\times$ lower objective across $\mathrm{SNR}$ levels, indicating a substantially better starting point. Panel \textit{(d)} confirms this trend, with a consistent $\mathbf{35.6}\%$--$\mathbf{40}\%$ reduction in initial objective.

\textit{2) Interpretation: Prior Quality vs. Solver Speed:}
The iteration gain is modest: $4.55\%$ on average, peaking at $9.1\%$ at $\mathrm{SNR}=5\,\mathrm{dB}$, with one negative case at $\mathrm{SNR}=-5\,\mathrm{dB}$ ($-1.6\%$). This suggests that the main benefit of the GNN is not faster convergence, but a better initialization. Cold-start and warm-start reach nearly identical final solutions, as reflected by the matching final objective values in Table~\ref{tab:event1_combined}, panel \textit{(c)}. Thus, the GNN mainly acts as a physics-consistent prior that denoises the initial estimate.

\subsection{Combined Performance Summary}
\label{sec:results_combined}

Table~\ref{tab:event1_combined} consolidates the three benchmarks into a single view for method selection. Panel \textit{(a)} compares integration methods: Forward Euler has the lowest cost ($1.06\,\mathrm{ms}$) but weaker convergence, whereas Trapezoidal offers a better balance ($1.30\,\mathrm{ms}$, about $22\%$ overhead) with $\sim 61\%$ convergence. Panel \textit{(b)} shows solver trade-offs: Levenberg--Marquardt delivers strong robustness ($95\%$ convergence) at a modest extra cost ($1.79\,\mathrm{ms}$, about $9\%$ overhead). Panel \textit{(c)} highlights the GNN warm-start, with an average iteration improvement of $\mathbf{+4.55}\%$ across $\mathrm{SNR}$ levels and a peak gain of $\mathbf{+9.1}\%$ at $\mathrm{SNR}=5\,\mathrm{dB}$.

\begin{table*}[t]
\centering
\renewcommand{\arraystretch}{1.0}
\caption{{\small\scshape Trade-off summary and GNN warm-start benchmark}}
\label{tab:event1_combined}
\scriptsize
\setlength{\tabcolsep}{4pt}

\resizebox{\textwidth}{!}{%
\begin{minipage}{\textwidth}
\centering

\begin{minipage}[t]{0.485\textwidth}
\centering
\textbf{(a) Integration methods}\\[0.4em]
\begin{tabular}{
@{}l
S[table-format=1.4]
S[table-format=1.4]
S[table-format=2.4]
S[table-format=1.4]
@{}}
\toprule
\textbf{Method} &
\multicolumn{1}{c}{\textbf{Avg Time (ms)}} &
\multicolumn{1}{c}{\textbf{Avg Residual}} &
\multicolumn{1}{c}{\textbf{Avg Conv. (\%)}} &
\multicolumn{1}{c}{\textbf{Avg Iters}} \\
\midrule
Euler Forward  & 1.0642 & 0.0649 & 65.8867 & 3.3333 \\
Trapezoidal    & 1.2956 & 0.0678 & 61.0698 & 4.1667 \\
Quadratic      & 1.5418 & 0.0699 & 61.1196 & 4.8333 \\
Backward Euler & 1.9125 & 0.0651 & 60.3538 & 5.6667 \\
\bottomrule
\end{tabular}
\end{minipage}
\hfill
\begin{minipage}[t]{0.485\textwidth}
\centering
\textbf{(b) Iterative solvers}\\[0.4em]
\begin{tabular}{
@{}l
S[table-format=1.4]
S[table-format=1.4]
S[table-format=2.4]
S[table-format=1.4]
@{}}
\toprule
\textbf{Solver} &
\multicolumn{1}{c}{\textbf{Avg Time (ms)}} &
\multicolumn{1}{c}{\textbf{Avg Residual}} &
\multicolumn{1}{c}{\textbf{Avg Conv. (\%)}} &
\multicolumn{1}{c}{\textbf{Avg Iters}} \\
\midrule
Levenberg--Marquardt & 1.7866 & 0.0538 & 95.1667 & 5.3333 \\
Hybrid GN$\rightarrow$LM & 1.6360 & 0.0556 & 95.6667 & 4.8333 \\
Gauss--Newton & 1.6529 & 0.0557 & 66.6667 & 4.3333 \\
\bottomrule
\end{tabular}
\end{minipage}

\vspace{1.0em}

\begin{minipage}[t]{\textwidth}
\centering
\textbf{(c) GNN warm-start vs.\ cold-start GN (Quadratic DSE)}\\[0.4em]
\setlength{\tabcolsep}{6pt}
\begin{tabular*}{\linewidth}{
@{\extracolsep{\fill}}
S[table-format=-2.0]
S[table-format=1.6]
S[table-format=1.6]
S[table-format=-1.6]
S[table-format=1.2e-1]
S[table-format=1.2e-1]
S[table-format=1.6]
@{}}
\toprule
\multicolumn{1}{c}{\textbf{SNR (dB)}} &
\multicolumn{1}{c}{\textbf{Cold Iters}} &
\multicolumn{1}{c}{\textbf{GNN Iters}} &
\multicolumn{1}{c}{\textbf{$\Delta$ Iters (\%)}} &
\multicolumn{1}{c}{\textbf{Cold $J$}} &
\multicolumn{1}{c}{\textbf{GNN $J$}} &
\multicolumn{1}{c}{\textbf{$\Delta J$ (\%)}} \\
\midrule
 20  & 3.141667 & 2.991667 &  4.774536 & 6.70e-5 & 6.70e-5 & 0.000000 \\
 15  & 3.133333 & 2.983333 &  4.787234 & 6.70e-5 & 6.70e-5 & 0.000000 \\
 10  & 3.150000 & 3.000000 &  4.761905 & 7.00e-5 & 7.00e-5 & 0.000000 \\
  5  & 3.300000 & 3.000000 &  9.090909 & 7.90e-5 & 7.90e-5 & 0.000000 \\
  0  & 3.191667 & 3.041667 &  4.699739 & 7.90e-5 & 7.90e-5 & 0.000000 \\
 -5  & 3.166667 & 3.216667 & -1.578947 & 1.28e-4 & 1.28e-4 & 0.000000 \\
-10  & 3.283333 & 3.108333 &  5.329949 & 1.95e-4 & 1.95e-4 & 0.000000 \\
\midrule
\multicolumn{7}{c}{\scriptsize Average iteration improvement (GNN vs.\ cold): $+4.55\%$; average objective improvement: $0.00\%$} \\
\bottomrule
\multicolumn{7}{c}{\scriptsize $\Delta(\%) = 100\cdot\left(1-\frac{\text{GNN}}{\text{Cold}}\right)$; negative means worse for GNN.}
\end{tabular*}
\end{minipage}

\end{minipage}%
}
\end{table*}

\section{Discussion}
\label{sec:discussion}

Three key observations emerge from the benchmarking results. First, clear trade-offs are observed between integration accuracy, computational cost, and convergence robustness. Forward Euler is the fastest method overall, but its convergence degrades substantially at low SNR, which limits its suitability in high-noise operating conditions. By contrast, the Trapezoidal method provides a more practical balance, achieving approximately $\mathrm{61}\%$ convergence with only about $\mathrm{22}\%$ additional computational overhead relative to the fastest baseline. Higher-order schemes such as Quadratic and Backward Euler offer improved numerical fidelity, but their additional cost makes them more appropriate for offline studies or applications in which estimation accuracy is prioritized over execution speed. Second, the solver comparison indicates that robustness under noisy conditions requires damping. Gauss--Newton achieves only about $\mathrm{67}\%$ convergence at low SNR because of frequent divergence, whereas Levenberg--Marquardt restores convergence to roughly $\mathrm{95}\%$ across all SNR levels. The hybrid GN$\rightarrow$LM strategy provides the most balanced performance, combining strong robustness of approximately $\mathrm{95.7}\%$ with relatively low computational cost, requiring only about $\mathrm{1.64}\,\mathrm{ms}$ and $\mathrm{4.8}$ iterations on average. This suggests that damping is essential for stable estimation in noisy CT scenarios, while hybrid switching offers a practical compromise between numerical reliability and speed. Third, the GNN warm-start results support the interpretation of learning as a regularization prior rather than a replacement for the physics-based solver. The learned initializer reduces the initialization distance by approximately $\mathrm{25}\%$, and this improvement persists even at very low SNR levels such as $-\mathrm{10}\,\mathrm{dB}$. This behavior indicates that the GNN captures a physics-consistent prior over the CT state, improving the initial point supplied to the iterative estimator. At the same time, the average iteration improvement remains modest at approximately $\mathrm{4.55}\%$, which suggests that the main benefit of the GNN lies in improved early optimization conditions rather than dramatically faster convergence. In practice, both cold-start and warm-start strategies converge to nearly identical final states, but the warm-started solution reaches the convergence region from a more favorable initialization.

\section{Conclusion}
\label{sec:conclusion}

This study investigated the combined effects of integration methods, iterative solvers, and learned initialization on dynamic state estimation for current-transformer saturation correction. Three key contributions emerge: (1) systematic benchmarking of four integration schemes and three solvers across $\mathrm{SNR}$ levels from $-10$ to $+20\,\mathrm{dB}$; (2) a physics-informed GNN initializer achieving $25\%$ initialization distance reduction and $39\%$ initial objective improvement; and (3) practical deployment guidance for algorithm selection under noisy conditions. Results demonstrate that iterative solver choice is critical for robustness, improving convergence from $67\%$ to $95\%$ at low $\mathrm{SNR}$ when damping is introduced. Integration method selection determines the computational-accuracy trade-off, with higher-order schemes incurring $48$--$79\%$ additional overhead. The GNN warm-start provides essential regularization in noisy settings by delivering a favorable initialization to the physics-based solver. Future work will pursue cross-event generalization on different saturation events with varying characteristics, mixed-$\mathrm{SNR}$ training for non-stationary noise environments, comparative evaluation against LSTM and ResNet baselines, real-time deployment on embedded hardware (ARM Cortex-M, FPGA), and adaptive solver switching via online $\mathrm{SNR}$ estimation. These directions will enhance practical applicability and establish a foundation for adaptive CT state estimation.

\bibliographystyle{IEEEtran}
\bibliography{6320}

\vspace{0.1cm}

\begin{IEEEbiography}[\vspace{-1.2cm}
{\includegraphics[width=1in,height=1.3in,clip,keepaspectratio]{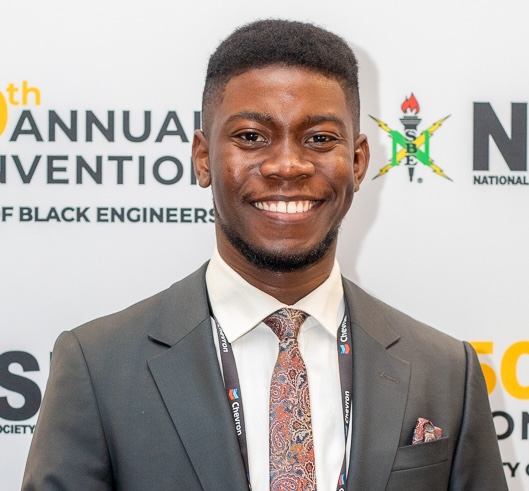}}]
{Michael Boateng}
(Student Member, IEEE) received the B.S. degree in electrical engineering from Ashesi University, Accra, Ghana in 2022. He is currently working toward the Ph.D. degree in electrical and computer with the Georgia Institute of Technology, Atlanta, GA, USA. His research focuses on optimization, machine learning, and power systems.
\end{IEEEbiography}

\vspace{-1.2cm} 

\begin{IEEEbiography}[\vspace{-1.2cm}
{\includegraphics[width=1in,height=1.3in,clip,keepaspectratio]{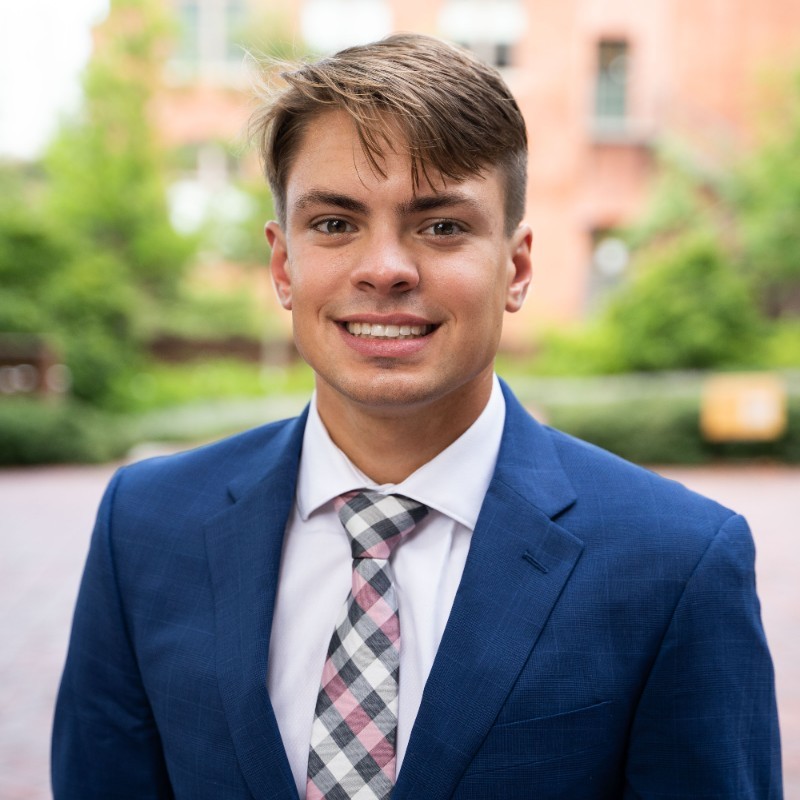}}]
{Gabe Gauderman}
(Student Member, IEEE) received the B.S. degree in electrical engineering from the Georgia Institute of Technology, in 2024. He completed the MS degree in electrical and computer engineering at the Georgia Institute of Technology, Atlanta, GA, USA in 2025. He is currently working as a distribution system planning and strategy engineer at Xcel Energy
\end{IEEEbiography}

\vspace{-1.2cm} 

\begin{IEEEbiography}[\vspace{-1.2cm}
{\includegraphics[width=1in,height=1.3in,clip,keepaspectratio]{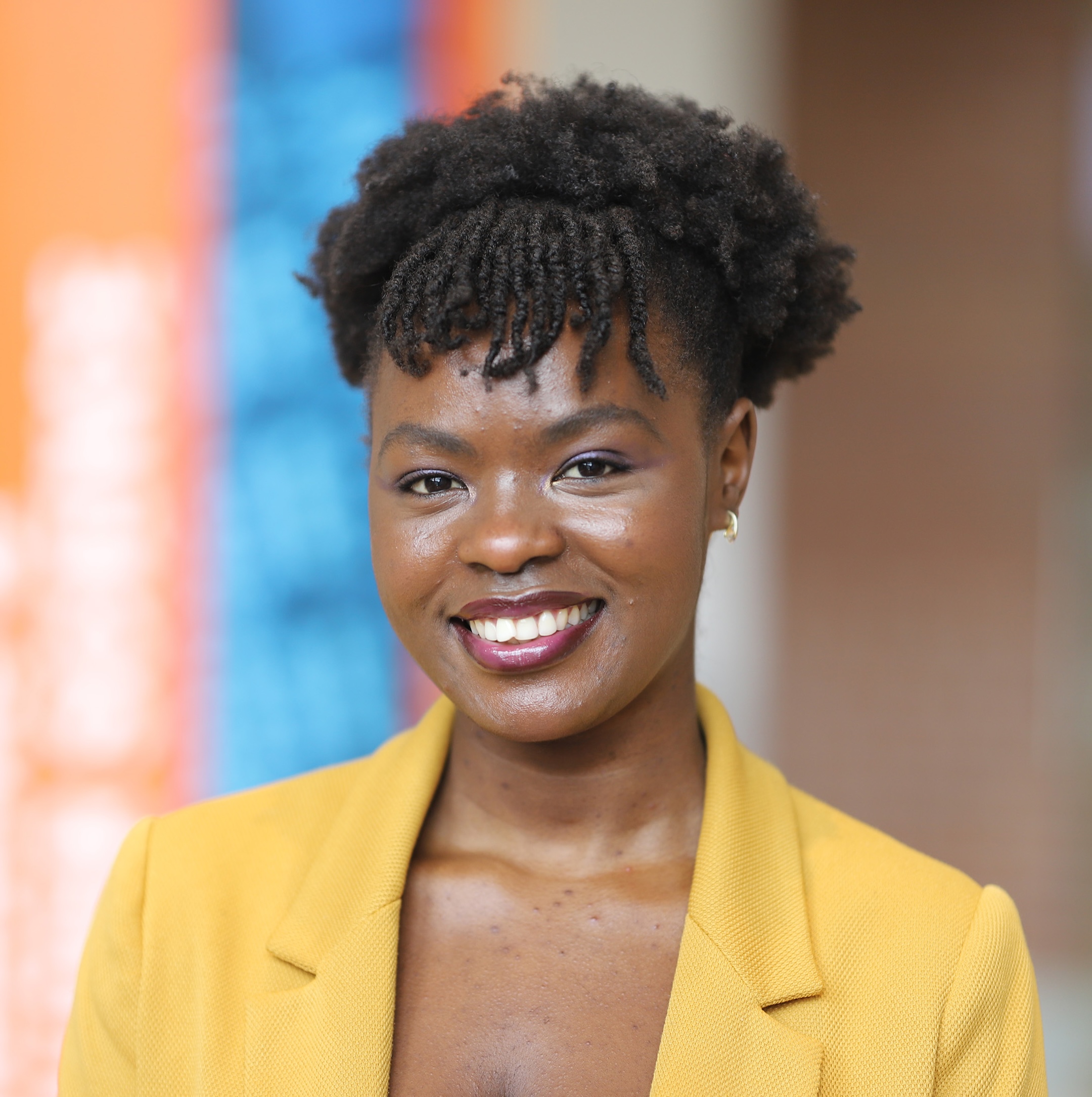}}]
{Nathalie Uwamahoro} (Student Member, IEEE) received the B.Sc. degree in electrical engineering from the University of Rwanda in 2017 and the M.Sc. degree in electrical and computer engineering from Carnegie Mellon University in 2019. Currently working toward the Ph.D. degree in electrical engineering at Syracuse University, USA. Her research focuses on machine learning and power systems.
\end{IEEEbiography}

\vfill

\appendix



\vspace{-0.5em}

\begin{table}[!h]
\centering
\renewcommand{\arraystretch}{1.0}
\caption{{\small\scshape CT Measurement Model for Dynamic State Estimation}}
\label{tab:measurement_model_appendix}
\scriptsize
\setlength{\tabcolsep}{2pt}

\resizebox{\columnwidth}{!}{%
\begin{tabular}{c p{6.5cm} c}
\toprule
\textbf{Index} & \textbf{Measurement function $z_i = h_i(\mathbf{x}_k,\mathbf{x}_{k-1})$} & \textbf{$\sigma$} \\
\midrule
1  & $z_1 = v_{1}(t)-v_{1}(t-1)$ & $0.005~\mathrm{V}$ \\

2  & $z_2 = \dfrac{1}{\tau}\!\left(i_{1}(t)-i_{1}(t-1)\right) + \dfrac{h_{1}(t)}{L_{11}}\!\left(i_{2}(t)-i_{2}(t-1)\right) + \dot{i}_{1}(t)$ & $0.005~\mathrm{A}$ \\

3  & $z_3 = \dfrac{1}{\tau}\!\left(i_{2}(t)-i_{2}(t-1)\right) + \dfrac{h_{2}(t)}{L_{12}}\!\left(i_{1}(t)-i_{1}(t-1)\right) + \dot{i}_{2}(t)$ & $0.005~\mathrm{A}$ \\

4  & $z_4 = g_{1}i_{1}(t)+g_{2}i_{2}(t)+\dot{i}_{2}(t)$ & $0.005~\mathrm{A}$ \\

5  & $z_5 = \dfrac{d}{dt}\!\left(x_{1}(t)-x_{2}(t)\right)$ & $0.0005~\mathrm{A}$ \\

6  & $z_6 = x_{2}(t)-x_{3}(t)+\dot{x}_{3}(t)$ & $0.0005~\mathrm{A}$ \\

7  & $z_7 = i_{1}(t)-i_{2}(t)$ & $0.0005~\mathrm{A}$ \\

8  & $z_8 = \dfrac{d}{dt}\!\left(i_{1}(t)-i_{2}(t)\right)$ & $0.0005~\mathrm{A}$ \\

9  & $z_9 = \dfrac{d}{dt}\!\left(v_{1}(t)-v_{2}(t)\right)$ & $0.0005~\mathrm{V}$ \\

10 & $z_{10} = \dfrac{d}{dt}\!\left(v_{2}(t)-v_{3}(t)\right)$ & $0.0005~\mathrm{V}$ \\

11 & $z_{11} = \dfrac{1}{\tau}\!\left(v_{1}(t)-v_{2}(t)\right) + \dfrac{1}{L_{22}}\!\left(v_{3}(t)-v_{2}(t)\right) + \dot{v}_{1}(t)$ & $0.0005~\mathrm{V}$ \\

12 & $z_{12} = \dfrac{d}{dt}\!\left(i_{3}(t)-i_{4}(t)\right)$ & $0.0005~\mathrm{A}$ \\

13 & $z_{13} = \dfrac{d}{dt}\!\left(v_{3}(t)-v_{4}(t)\right)$ & $0.0005~\mathrm{V}$ \\

14 & $z_{14} = \dfrac{d}{dt}\!\left(i_{4}(t)-i_{5}(t)\right)$ & $0.0005~\mathrm{A}$ \\

15 & $z_{15} = g_{3}i_{3}(t)+g_{4}i_{4}(t)+\dot{i}_{5}(t)$ & $0.0005~\mathrm{A}$ \\

16 & $z_{16} = g_{4}i_{4}(t)+\dot{i}_{6}(t)$ & $0.0005~\mathrm{A}$ \\

17 & $z_{17} = \dfrac{d}{dt}\!\left(x_{4}(t)-x_{5}(t)\right)$ & $0.0005~\mathrm{A}$ \\

18 & $z_{18} = g_{5}i_{5}(t)+\dfrac{d}{dt}\!\left(i_{6}(t)\right)$ & $0.005~\mathrm{A}$ \\

19 & $z_{19} = \dfrac{d}{dt}\!\left(x_{5}(t)-x_{6}(t)\right)$ & $0.0005~\mathrm{A}$ \\

20 & $z_{20} = \dfrac{d}{dt}\!\left(x_{6}(t)\right)$ & $0.0005~\mathrm{A}$ \\

21 & $z_{21} = \tau g_{s}\!\left(\theta(t)-\theta(t-1)\right)$ & $0.05~\mathrm{A}$ \\
\bottomrule
\end{tabular}%
}
\end{table}

\end{document}